\documentclass[aip,jcp,reprint]{revtex4-1} 
\usepackage{graphicx}
\usepackage{enumitem}

\usepackage{amsmath,amssymb}

\usepackage{lipsum}

\usepackage{mathtools}
\usepackage{graphicx}   
\usepackage{color}
\usepackage{ulem}
\usepackage{caption}
\usepackage{subcaption}
\usepackage{comment}
\usepackage{bm}

\usepackage{array}
\usepackage{booktabs}
\usepackage{multirow}

\usepackage{amsmath}

\usepackage{booktabs}

\DeclareMathOperator\erf{erf}
\DeclareMathOperator\erfc{erfc}

\begin{document}
\title{Influence of long range forces on the transition states and dynamics of NaCl ion-pair dissociation in water}

	\author{Dedi Wang}
		 \thanks{These two authors contributed equally.}
 \affiliation{Biophysics Program and Institute for Physical Science and Technology,
 University of Maryland, College Park 20742, USA.}
 
 \author{Renjie Zhao}
 	 \thanks{These two authors contributed equally.}
 \affiliation{Chemical Physics Program and Institute for Physical Science and Technology,
 University of Maryland, College Park 20742, USA.}
 
 \author{John D. Weeks\footnote{Corresponding author.}}
 \email{jdw@umd.edu}

 \affiliation{Department of Chemistry and Biochemistry and Institute for Physical Science and Technology,
 University of Maryland, College Park 20742, USA.}

 \author{Pratyush Tiwary\footnote{Corresponding author.}}
 
 \email{ptiwary@umd.edu}
 \affiliation{Department of Chemistry and Biochemistry and Institute for Physical Science and Technology,
 University of Maryland, College Park 20742, USA.}

	\date{\today}
	
\begin{abstract}

We study NaCl ion-pair dissociation in a dilute aqueous solution using computer simulations both for the full system with long range Coulomb interactions and for a well chosen reference system with short range intermolecular interactions.
Analyzing results using concepts from Local Molecular Field (LMF) theory and the recently proposed AI-based analysis tool ``State predictive information bottleneck" (SPIB) we show that the system with short range interactions can accurately reproduce the transition rate for the dissociation process, the dynamics for moving between the underlying metastable states, and the transition state ensemble. Contributions from long range interactions can be largely neglected for these processes because long range forces from the direct interionic Coulomb interactions are almost completely canceled ($>90\%$) by those from solvent interactions over the length scale where the transition takes place.
Thus for this important monovalent ion-pair system, short range forces alone are able to capture detailed consequences of the collective solvent motion, allowing the use of physically suggestive and computationally efficient short range models for the disassociation event. We believe that the framework here should be applicable to disentangling mechanisms for more complex processes such as multivalent ion disassociation, where previous work has suggested that long range contributions may be more important.   
\end{abstract}

	\maketitle
\section{Introduction}

A wide variety of physical and biological processes such as protein folding,\cite{levy2006water,dill1990dominant} self-assembly of dissolved molecules\cite{venkateshwaran2014water} and nucleation at interfaces,\cite{fang2016nucleation} take place in aqueous environments.
 A concise and accurate description of these systems requires an efficient treatment of the interplay between the short and long range components of the intermolecular interactions reflected in the local hydrogen bond network in water and dielectric screening of distant solute charges.


Most computer simulations use periodic boundary conditions to address these issues, where contributions from long range Coulomb interactions in periodic images are determined by Ewald and related lattice sum methods.\cite{allen2017computer,belhadj1991molecular,essmann1995smooth} However, this can introduce computation overhead due to their relatively poor scaling with the system size\cite{schulz2009scaling} and to
parallel scaling limitations from communication latency.\cite{sun2012optimizing,kutzner2014scaling} Moreover, physical insight into the different roles of short and long range interactions in equilibrium and dynamic properties is often obscured by the indirect “black-box” nature of the lattice sum algorithms.

Indeed, the classic problem of ion-pair dissociation of NaCl in water has long presented challenges  to many simple pictures of solvent-mediated processes.\cite{Smith1994,Geissler1999,Ballard2012,Mullen2014} Work by Geissler, Dellago, and Chandler first demonstrated that the ion-pair distance $r_{ion}$ alone is not a sufficient reaction coordinate (RC) to describe the dissociation dynamics, and reorganization of the surrounding solvent must be taken into account.\cite{Geissler1999} Ballard and Dellago showed that the dissociation event can be affected by solvent perturbations extending out into the third solvation shell.\cite{Ballard2012} However, a latter paper by Peters and coworkers \cite{Mullen2014} argued that information only from water molecules in the first solvation shell was needed to  predict the committor, which is arguably a perfect candidate for the RC and determines key features of the solvation dynamics, such as the transition state ensemble.\cite{committor_review}

In this paper we revisit this  problem using some recent methods that we believe can be naturally extended to more complicated solvent-mediated processes such as protein conformational changes and crystal nucleation. 
Rather than characterizing events in terms of solvent shell correlations, our focus here is to determine the extent to which simplified and computationally efficient reference models with well chosen short range intermolecular interactions can give an accurate description of relevant static and dynamic processes in the full long range system. 
As detailed in Sec.\ III, our results show that short range interactions alone can reproduce most of the previously determined dynamic properties of the NaCl ion-pair dissociation process including the existence of solvent perturbations extending to the third solvation shell and many features of the transition state ensemble.

Specifically, we employ a combination of Local Molecular Field (LMF) theory\cite{chen2004connecting,chen2006local,rodgers2008interplay,rodgers2008local,weeks2002connecting,weeks1971role} and our recently proposed artificial intelligence (AI)-based analysis tool ``State predictive information bottleneck" (SPIB).\cite{SPIB}
The LMF treatment of aqueous solvated systems uses a minimal reference model for disentangling the different effects of short and long range interactions on equilibrium structural and thermodynamic properties, exploiting the near cancellation of slowly varying long range Coulomb forces in uniform environments. In previous work,\cite{remsing2011deconstructing} the short range Gaussian truncated (GT) water model, discussed in Sec. II below, was shown to accurately reproduce atom-atom correlation functions of the bulk water solvent and many properties of the hydrogen bond network, and it further revealed the dominant role of short range interactions in the anomalous behavior of the internal pressure and the density maximum.

Recently an LMF-based short solvent model was developed for predicting the potential of mean force (PMF) between charged solutes in dilute aqueous solutions.\cite{gao2020short}
The short solvent model employs a reference system where all Coulomb interactions involving solvent molecules are Gaussian truncated and modified long range interactions are taken into account only between solute charges. For the dilute NaCl solution of interest here, it was previously found that these renormalized ion-ion corrections could be neglected as well, with little deviation from the cation-anion PMF of the full system.\cite{gao2020short} Thus a minimal short range model where all Coulomb interactions are Gaussian truncated is sufficient to describe the equilibrium PMF.

Here we investigate the extent to which the minimal model can also recover the dissociation dynamics of NaCl by analyzing high-dimensional trajectories generated from molecular dynamics (MD) simulations of full and short range systems using insights from the SPIB method.
SPIB approximates the RC as a past-future information bottleneck, defined as the minimal amount of information needed to be known about a system to accurately predict its future  state. 

The SPIB approach has a number of useful features. First, it can automatically partition the high-dimensional state space into a few metastable states with minimal human intuition regarding the number or location of metastable states, and also analyze the dynamics governing state-to-state movement. Second, it can identify the correct transition state ensembles just given a limited number of transitions. Third, inspired by Ref. \onlinecite{prave}, it provides the use of a tunable hyper-parameter called the time-delay $\Delta t$, which governs how far into the future predictions are being made. By gradually increasing $\Delta t$ from 0, one can filter out all the fast modes and control the level of coarse-graining of the system.

Through this combination of LMF and SPIB, we can carry out a systematic kinetic analyses of NaCl ion-pair dissociation in water. We first study the free energy barrier and the transition rate for the dissociation process. Then we use SPIB to identify underlying metastable and transition states for the dissociation process. The transitions between these metastable states and the position of the transition state ensemble obtained by SPIB are then carefully analyzed and compared to full system results.


\section{Methods}
\label{sec:Methods}

\subsection{LMF theory and the Complete Gaussian Truncated solvent model}

Building on ideas originally used for LJ potentials, \cite{weeks1971role,chandler1983van} LMF theory asserts that Coulomb intermolecular potentials in the full system of interest can be usefully separated into strong short range and uniformly slowly-varying long range components. To that end, we rewrite the potential $v(r)$ from a unit point charge as
\begin{equation}
v(r) = \frac{1}{r}=\frac{\erfc{(r/\bar{\sigma})}}{r}+\frac{\erf{(r/\bar{\sigma})}}{r}=v_0(r)+v_1(r),
\end{equation}
where erf and erfc are the error and complementary error functions and  $\bar{\sigma}$ is a smoothing or truncation length usefully chosen to be a typical nearest neighbor distance in the full system.

The slowly varying component $v_1(r)$ in Eq.(1) is the potential of a unit Gaussian change distribution whose width is proportional to $\bar{\sigma}$.  $v_1(r)$ approaches $1/r$ at  large distances and remains slowly varying for $r$ less than $\bar{\sigma}$ where strong intermolecular core forces exist. By construction, $v_0(r)$ then accounts for the diverging short range Coulomb force at distances less than $\bar{\sigma}$ and quickly vanishes for distances larger than  $\bar{\sigma}$.

Applying these ideas to charges in the SPC/E water model used for the solvent in this paper, we define a \textit{Gaussian Truncated} (GT) version  by replacing the point charge potentials by the short range $v_0(r)$, while retaining the full oxygen LJ potential. 
Previous work has shown that GT water with any choice of  $\bar{\sigma}$ larger than the typical nearest-neighbor hydrogen bond distances in water (roughly 3.5~\AA) accurately reproduces the atom-atom correlation functions of bulk SPC/E water along with many detailed features of the local hydrogen bond network. \cite{remsing2011deconstructing}

To clarify the role of short range forces for ion-pair systems solvated in SPC/E water, we consider a simplified version of the short solvent model discussed in the introduction where all Coulomb interactions in the system are Gaussian truncated, including those between the ion pair charges. To determine an appropriate $\bar{\sigma}$ for this  minimal ``Complete Gaussian truncated" (CGT) solvent model, we note that the main influence of ions on the solvation structure is in the first hydration shell. In particular,  the Cl$^-$ anion can also form hydrogen bonds with water molecules through the balance between the strong short range electrostatic attraction and the repulsion involving its somewhat larger LJ core. Taking this into account by making a minimal choice of $\bar{\sigma}=5.0$~\AA, we would expect both water-water and ion-water hydrogen bonds to be well described by the CGT model.\cite{remsing2018water} 

Indeed, results using this CGT approximation in Ref. \onlinecite{gao2020short} showed that long range interactions in the NaCl system only negligibly affected the equilibrium PMF. It is thus natural to ask if a similar situation  would persist for the corresponding dynamical process of ion-pair dissociation. In this work, this is analyzed by comparing the dynamical behavior of the CGT model, analyzed by SPIB, to that exhibited by the full model, where Coulomb interactions are treated with conventional lattice sum methods.

\subsection{State Predictive Information Bottleneck (SPIB)}
State predictive information bottleneck (SPIB) has been described in detail in the original proof-of-principle publication\cite{SPIB} and here we summarize its key features. A more complete introduction to SPIB and its implementation details are discussed in Supplementary Material (SM). Similar to the existing RAVE family of methods,\cite{rave,prave} SPIB aims to uncover a low dimensional manifold (parameterized by a reaction coordinate or RC) on which the dynamics of the system can be projected. We take $\bm{X}$ as a high-dimensional signal characterizing a generic molecular system, and $\bm{y}$ as its corresponding state label, which is drawn from a dictionary of indices for possible metastable states. For the ion-pair dissociation problem considered here, $\bm{X}$ could be expressed in terms of some order parameters, such as the ion-pair distance, first and second shell coordination number and water density, while $\bm{y}$ could comprise the contact state and dissociated state. We describe these in detail in Sec. \ref{sec:solventcoord}. Let the values measured at time $t$ be denoted by $\bm{X}_t$ and $\bm{y}_t$ respectively. Then we assume that a good RC should carry the maximal predictive power for the future state of the system. In other words, a desired RC should use minimal information from the past signal $\bm{X}_t$ to predict its future state label $\bm{y}_{t+\Delta t}$ accurately. Such a learning process can be easily implemented through the deep variational information bottleneck framework.\cite{variational_IB} 

The workflow of SPIB is summarized as follows: First, we generate an initial guess of the state labels. For ion-pair dissociation, this can be achieved by simply dividing the input data space into a sufficiently fine grids along the ion-pair distance as our initial state labels. We then input the trajectory data $\bm{X}$ and the state labels $\bm{y}$ into SPIB, and try to find the optimum RC which captures the most important features of the past configuration $\bm{X}_t$ to predict the future state $\bm{y}_{t+\Delta t}$. After this learning process, we can refine the state labels based on the learned RC, and the new refined state labels are then fed back into SPIB. The processes will be repeated until the converged RC and state labels are generated for further analyses. 

\subsection{Simulation Methods}

In our simulations, we consider a system composed of 727 SPC/E water molecules and the NaCl ion-pair. Following Ref. \onlinecite{gao2020short}, the ions are modeled as point charges embedded in LJ cores with parameters  $\epsilon_{Na}=\epsilon_{Cl}=0.1$ kcal/mol, $\sigma_{Na}=2.586$~\AA~and $\sigma_{Cl}=4.404$~\AA, while the O$^-$ anion of the SPC/E water has LJ parameters $\epsilon_{O}=0.15535$ kcal/mol and $\sigma_{O}=3.166$~\AA .
Lorentz–Berthelot mixing rules \cite{lorentz1881ueber} are used to determine the cross-interaction parameters. Both the CGT model and the full model are simulated using the LAMMPS package. \cite{plimpton1995fast}

The simulations are performed with a time step of 1 fs in the isothermal isobaric (NPT) ensemble at $T = 300$ K and $P = 1$ atm, realized by a Nose-Hoover thermostat\cite{nose1984unified} and barostat\cite{nose1983constant} with damping constants of $0.1$ and $1.0$ ps, respectively. For each system, a 100 ns long equilibrium trajectory is generated for further kinetic analyses. A total of 74 dissociation events through multiple transition pathways are observed in the full system, while a total of 55 dissociation events are observed in the CGT system.
Further discussion of the two models is given in the SM.
 
\section{Results}
\label{sec:Results}

\subsection{Solvent Coordinates}
\label{sec:solventcoord}
Previous work has established that the collective motion of the solvent plays an important role in the NaCl ion-pair dissociation.\cite{Geissler1999,Ballard2012,Mullen2014} Thus, by referring to Ref. \onlinecite{Mullen2014}, in addition to the distance between ions ($r_{ion}=|\bm{r}_{cation}-\bm{r}_{anion}|$), we selected 12 other order parameters (OPs) to account for details of the solvent arrangements. The definition of all the 13 OPs considered is provided in the SM. All of these OPs depend only on the instantaneous local configuration of solvent molecules, such as the coordination number up to the second solvation shell. Among them, the number of bridging waters $N_B$ and the interionic water density $\rho_{i3}$ are used for projection and visualization in the main text to remain consistent with Ref. \onlinecite{Mullen2014}.  After rescaling all the values into a range of $[0,1]$, the simulation trajectory data in terms of these 13 OPs was used as the input $\bm{X}$ of SPIB to analyze the kinetics of the NaCl ion-pair system. 

\subsection{Fundamental Kinetic Analysis}
\begin{figure}[ht]
    \centering
    \includegraphics[width=0.45\textwidth]{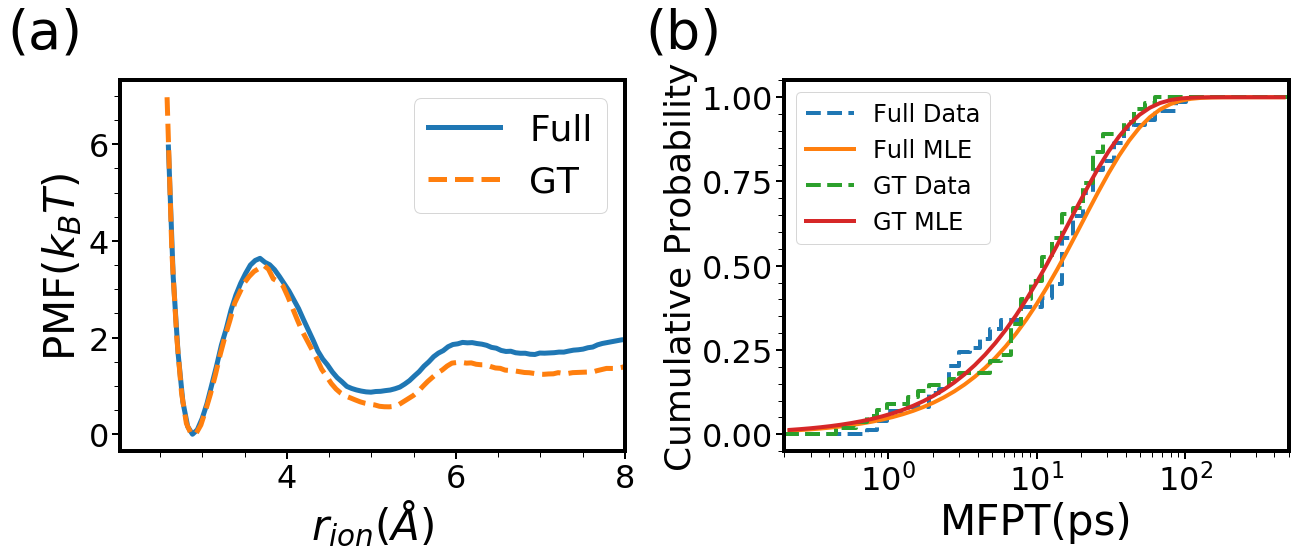}
    \caption{(a) The potential of mean force (PMF) of the aqueous NaCl system as a function of ion-pair distance $r_{ion}$. The PMF is calculated as $-k_{B}T\log{g(r_{ion})}$, where $g(r)$ is the radial distribution function. The local maximum of PMF at $r_{ion}=3.7$~\AA~ separates the contact state ($r_{ion}<2.9$~\AA) and dissociated state ($r_{ion}>5.0$~\AA). (b) The mean first-passage time (MFPT) from the contact state ($r_{ion}<2.9$~\AA) to the dissociated state ($r_{ion}>5.0$~\AA~) for different systems. The dashed lines denote the histogram data, while solid lines show corresponding best fits from maximum likelihood estimation (MLE).}
    \label{fig:fundamental_analysis}
\end{figure}

The cation–anion PMF calculated from the long unbiased simulation is shown in Fig. \ref{fig:fundamental_analysis}(a). Though at least 3 basins can be seen along the distance between ions ($r_{ion}$) based on the PMF, most workers have focused  on the largest barrier to dissociation.\cite{Smith1994,Geissler1999,Ballard2012,Mullen2014} Thus, the barrier peak at about $r_{ion} = 3.7$~\AA~ and $3.6~k_{B}T$ is typically used to distinguish the contact ion-pair from the solvent-separated ion-pair. The PMFs obtained here are consistent with  previous results from umbrella sampling\cite{gao2020short} with no significant differences in the PMFs, especially in the first barrier to dissociation in 
the full system and CGT system. 

A further kinetic analysis of the transition rate also confirms this finding. Based on the basin positions shown in Fig. \ref{fig:fundamental_analysis}(a), we define the contact state as $r_{ion}<2.9$~\AA~ and the solvent-separated state as $r_{ion}>5.0$~\AA. Fig. \ref{fig:fundamental_analysis}(b) shows the distribution functions and their best-fit Poisson curves of the mean first-passage time (MFPT) from the contact state to the dissociated state for different systems. Maximum likelihood estimation (MLE) is used to estimate the mean first-passage time and gives $\tau=20\pm2\ ps$ for the full system and $\tau=17\pm2\ ps$ for the CGT system. Then we test the goodness-of-fit based on the Kolmogorov–Smirnov (KS) statistic,\cite{Salvalaglio2014} which gives $p\text{-value}=0.57$ and $0.62$, respectively. This test result suggests that the mean first-passage time does follow a Poisson distribution and that our state definition seems reasonable. The transition time for the full system also agrees with the previous results from transition path sampling and reactive flux calculations.\cite{Geissler1999}

From this kinetic analysis, it seems plausible that the CGT system with all Coulomb interactions truncated can reproduce basic features of the dissociation kinetics of the NaCl ion-pair in water. Moreover, by leveraging the power of SPIB, we can investigate the underlying dynamics in more detail.

\subsection{Metastable States Analysis}
\begin{figure}[ht]
    \centering
    \includegraphics[width=0.45\textwidth]{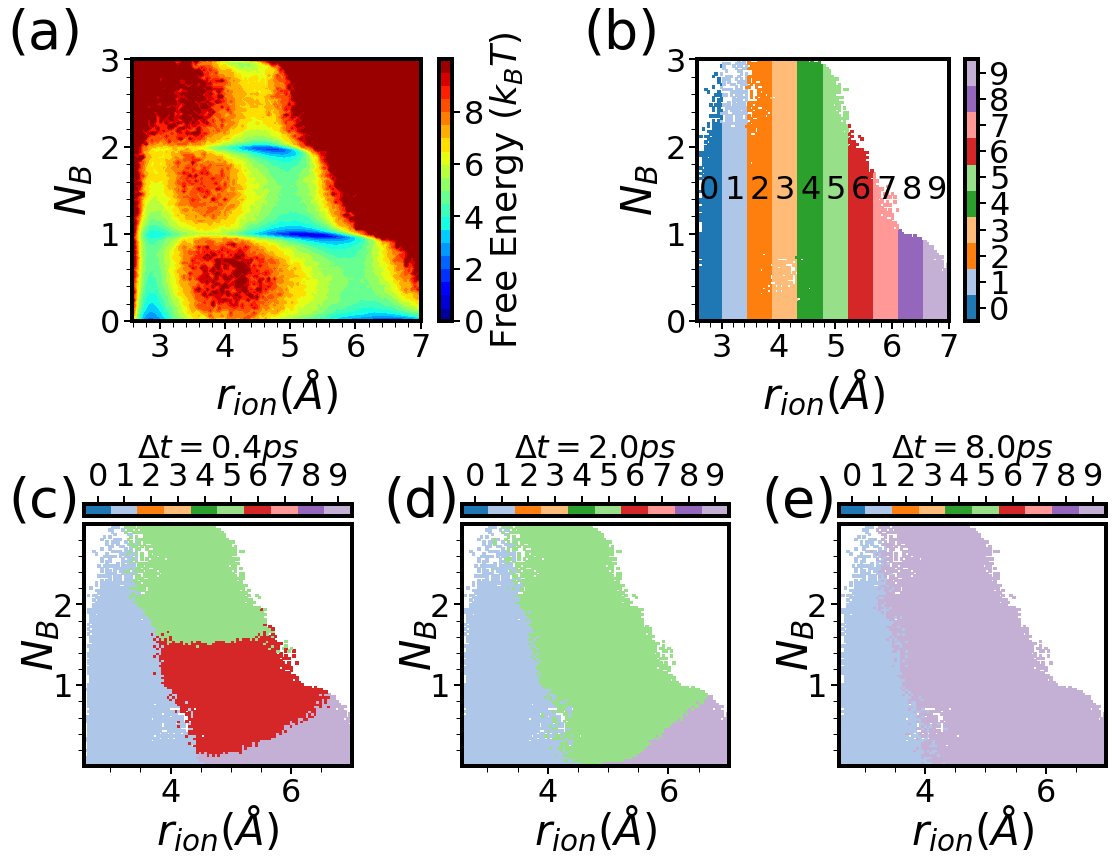}
    \caption{The free energy surface (a) and the time delay dependent discrete-state representation (b-e) for the full system projected onto coordinates $r_{ion}$ and $N_B$. The initial state labels are shown in (b), while the converged the state labels for different time delays are presented in (c-e). The state labels are learned using the time delay $\Delta t=0.4ps$ (c), $\Delta t=2.0ps$ (d), and $\Delta t=8.0ps$ (e) respectively. The color (or state label) in each grid corresponds only to the state label with highest fraction of samples for the respective grid point. }
    \label{fig:time_dependent_results}
\end{figure}

We now apply SPIB to the aqueous NaCl system where we do not assume any prior knowledge about the system such as the number and location of metastable states. In addition to $r_{ion}$, we introduce one solvent coordinate, the number of bridging waters $N_B$, to better visualize the system as shown in Fig. \ref{fig:time_dependent_results}(a). In this case, we arbitrarily divide the input data space into a fine grid used as our initial state labels (Fig. \ref{fig:time_dependent_results}(b)).

\begin{figure}[ht]
    \centering
    \includegraphics[width=0.45\textwidth]{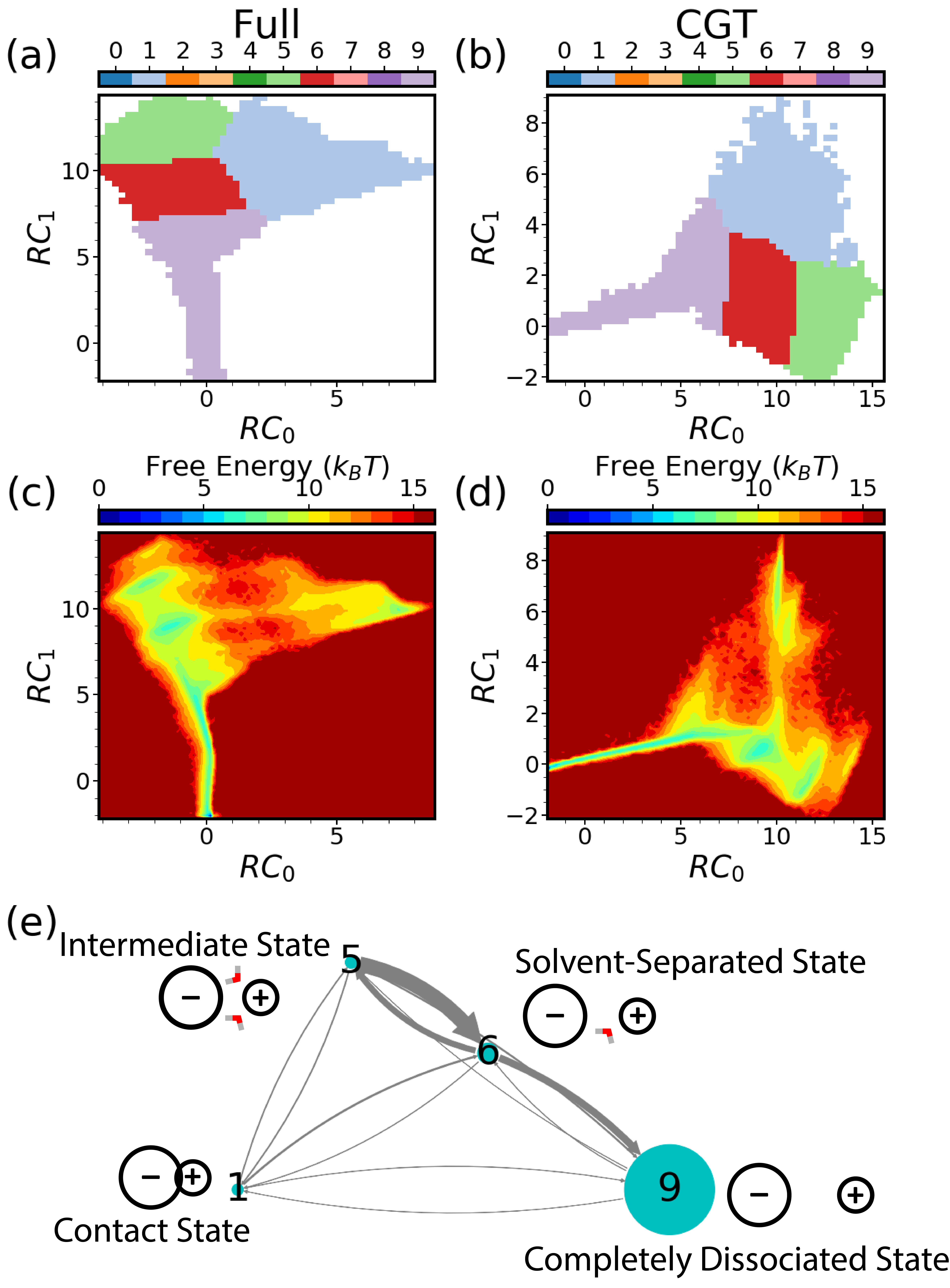}
    \caption{The multi-states dynamics learned by SPIB with the time delay $\Delta t=0.4ps$ for the full system (a,c) and GT system (b,d). The top row (a-b) are the state labels learned in the 2D RC space. The middle row (c-d) shows the free energy surface (-$k_BT\log P(RC_0,RC_1)$) in the 2D RC space. The bottom row (e) is the network representation of the transition probability matrix for the four-state kinetics. The final converged state 1 corresponds to the contact state, state 5 corresponds to the intermediate state, state 6 corresponds to the solvent-separated state, and state 9 corresponds to the completely dissociated state. The bridging water is oriented with a hydrogen toward Cl$^-$ and the oxygen toward Na$^+$ to compensate the attraction force between the ion-pair in the intermediate state and the solvent-separated state. The thickness of the arrows represents the transition probability.}
    \label{fig:four_states_dynamics}
\end{figure}

Fig. \ref{fig:time_dependent_results} reveals the underlying complexity in even such an apparently simple system. In Fig. \ref{fig:time_dependent_results}(c), we show how SPIB with a small time delay can automatically identify the metastable states in this system (numbered state 1, state 5, state 6 and state 9 by SPIB). The meaning of these metastable states is summarized in Fig. \ref{fig:four_states_dynamics}(e). In addition to the contact state (state 1), the solvent-separated state (state 6) and the completely dissociated state (state 9) shown on the profile of the potential of mean force (Fig. \ref{fig:fundamental_analysis}(a)), we can identify another metastable state (numbered state 5). 
The existence of this metastable state was discussed in Ref. \onlinecite{Mullen2014} as a possible intermediate state that could play an important role in explaining the recrossing behaviour of the ion-pair dissociation. 

In fact, two additional intermediate states can be identified by SPIB using an even smaller time delay ($\Delta t=0.1~ps$, see SM), but they have much shorter lifetimes, so we will focus on the intermediate state 5 shown here which has a much longer lifetime (about $1~ps$). We will still refer to this newly identified state as an intermediate state in accordance with Ref. \onlinecite{Mullen2014}, though its state population and escape time are in fact comparable to those of the solvent-separated state (see SM). This intermediate state has very large overlap with the solvent-separated state in the projection of $r_{ion}$, so it cannot be identified solely from the 1D PMF in Fig. \ref{fig:fundamental_analysis}(a). 
But a clear correspondence can be observed between the four metastable states uncovered by SPIB shown in Fig. \ref{fig:time_dependent_results}(c) and the free energy minima in the $r_{ion}$-$N_B$ space shown in Fig. \ref{fig:time_dependent_results}(a). A more coarse-grained representation of the ion-pair system can be obtained by controlling the time delay $\Delta t$ as shown in Fig. \ref{fig:time_dependent_results}(c-e). Here only the results for the full system are presented, but the same time-dependent results are obtained for the CGT system (see SM).

\subsection{Interpreting the SPIB derived RC and constructing a Markov model}

Since SPIB can filter out all the fast processes and retain only the relevant slow dynamics of interest, here we can further compare the slow dynamics learned by SPIB for the full interaction system and the CGT system (Fig. \ref{fig:four_states_dynamics}). 
Fig. \ref{fig:four_states_dynamics}(a-d) indicate that similar 2D RCs can be learned by SPIB for these two systems with the time delay $\Delta t=0.4ps$. Such a 2D RC can provide detailed kinetic information about the metastable state dynamics. For instance, the intermediate state (state 5), the solvent-separated state (state 6) and the dissociated state (state 9) are close to each other and separated by very low barriers as shown in Fig. \ref{fig:four_states_dynamics}(c-d), while much higher barriers can be observed between the contact state (state 1) and them. This suggests the different interconversion timescales between these four metastable states. Moreover, the 2D RC also clearly shows that there is no direct transition between the intermediate state (state 5) and the dissociated state (state 9), while a direct transition between the contact state (state 1) and the dissociated state (state 9) exists.

To better interpret our RC, we adopt a method to identify the most important input order parameters by sequential randomization of input parameters. \cite{Kemp2007,Hummer2019} We have provided details of the procedure in SM. The top two important order parameters are the ion-pair distance $r_{ion}$ and the number of bridging water molecules $N_B$, which are used to visualize the different metastable states learned by SPIB in Fig. \ref{fig:time_dependent_results}. We also find that the learned RCs shared the same relevant components for both the full and CGT systems. Thus, given the similar RCs learned by SPIB, we are confident that removing all the long range interactions will not significantly affect the underlying dynamics. The implied timescale analysis further confirms that given the learned state representations, only small differences are observed in the relaxation timescales between the two systems (see SM). 

Based on the learned state representation, a four state Markov model can be built and the major kinetic pathways to dissociation are depicted in Fig. \ref{fig:four_states_dynamics}(e). 
Initially, bulk waters insert into the first solvation shell of Na$^+$ and Cl$^-$ and increase the total number of water molecules to 10 or 11 in preparation for the transition (see SM). At the same time, one or two water molecules orient to compensate the direct attractive force between the ion-pair. The fluctuation along $r_{ion}$ allows a transition to the solvent-separated state directly or via an intermediate state. The solvent-separated state is also unstable and quickly transitions to the completely dissociated state. 

We found all the transitions are accompanied by the insertion of bulk water into the first solvation shell, which leads to a total increase of 3 or 4 water molecules during the whole dissociation process. This finding agrees with the previous report for this system.\cite{Ballard2012} As mentioned earlier, even in the direct transition from the contact state to the solvent-separated state or the completely dissociated state, there exist some transient intermediate states. (See SM)   

\subsection{Transition State Ensemble Analysis}
Building on the analysis of the metastable states for both the full and CGT systems, we can use SPIB to further compare their transition state ensembles (TSEs) for the ion-pair dissociation process. However, the existence of the long-lived intermediate state identified in the last subsection makes it unreasonable to study only the direct transitions between the contact state and the solvent-separated state while ignoring all the intermediate states between them. 
Thus, we continue to take a four-state representation with a relatively small time delay ($\Delta t=0.4~ps$) as shown in Fig.\ref{fig:time_dependent_results}(c) to describe the dissociation process and define the transition state ensemble as the set of configurations whose transition probability to the contact state is $0.5$. 

\begin{figure}[ht]
    \centering
    \includegraphics[width=0.45\textwidth]{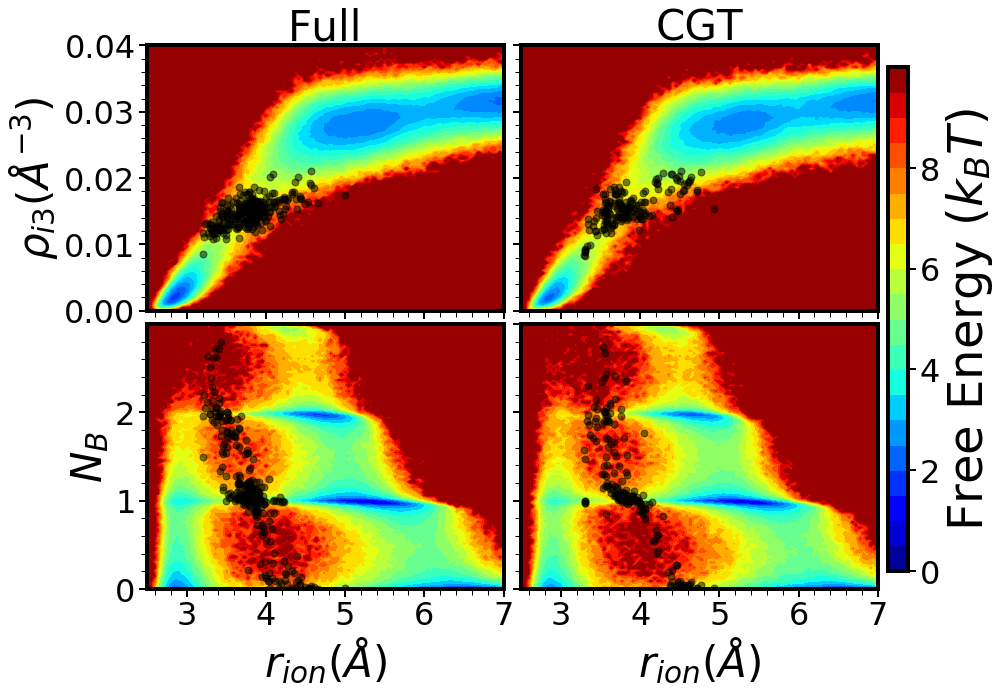}
    \caption{Free energy projected onto coordinates $r_{ion}$ and $\rho_{i3}$ (top row), $r_{ion}$ and $N_B$ (bottom row) for different systems. Transition state ensembles identified by SPIB with the time delay $\Delta t=0.4ps$ are projected onto each surface as black points.}
    \label{fig:TSE_results}
\end{figure}

Fig. \ref{fig:TSE_results} shows the projection of the transition state ensembles identified by SPIB onto the coordinates $r_{ion}$, $\rho_{i3}$ and $N_B$. The identified transition states in both the full and CGT systems are similar to the results of Ref. \onlinecite{Mullen2014}.
First, the higher the number of well-oriented interionic waters $N_B$  that cancel the the direct interionic force, the smaller is the $r_{ion}$ required to be in the transition states.
Second, all the transition states share an almost constant interionic water density $\rho_{i3}$, as shown in the top panel in Fig. \ref{fig:TSE_results}. To better interpret the TSE obtained by SPIB, we extract all the samples near the transition states ($0.01$~\AA$^{-3}<\rho_{i3}<0.018$~\AA$^{-3}$) for the input relevance analysis (see SM). The top three relevant order parameters are the number of bridging water molecules $N_B$, the interionic water density 
$\rho_{i4}$, which contains information similar to $\rho_{i3}$,
and the ion-pair distance $r_{ion}$, consistent with the findings of Ref. \onlinecite{Mullen2014}. Since the top 3 OPs in both systems are the same, this suggests that similar transition state ensembles would be identified by SPIB. We provide further details of the TSE analysis in SM.
\subsection{Explaining why long range forces have little effect on NaCl dynamics}
Thus far, we have shown that the short range interactions alone can reproduce many dynamical properties of NaCl ion-pair dissociation, including the transition rate, the metastable state kinetics, and the transition state ensemble. We now provide a mechanistic explanation for this peculiar finding.

To make a more detailed analysis, we measure their influence through the average solvent force on the solute.\cite{Geissler1999,Rey1992} We constrain the ion-pair distance $r_{ion}$ ranging from $3.25$ to $4.45$~\AA~ with a separation increment of $0.1$ \AA~ to obtain 13 constrained ensembles for both the full and the CGT systems. Each constrained trajectory is 1 ns long, and configurations are saved every 10 fs. We then use the trained SPIB to identify the transition state ensemble (TSE) from the saved configurations. 

\begin{figure}[ht]
    \centering
    \includegraphics[width=0.32\textwidth]{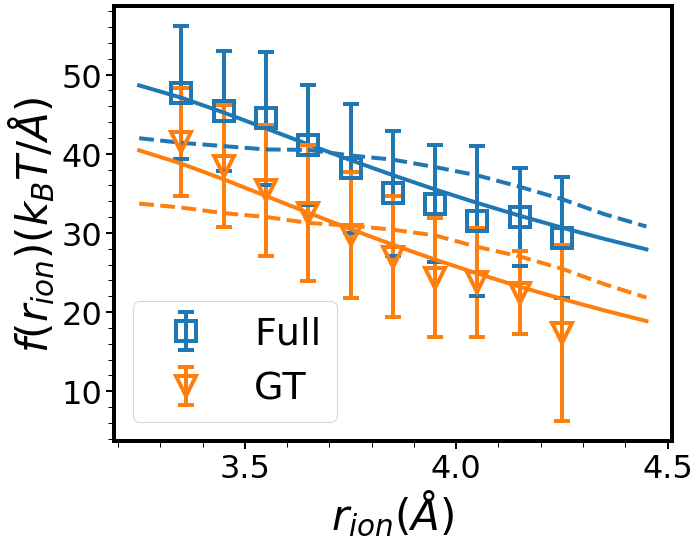}
    \caption{Mean solvent force on the ion-pair along the interionic axis as a function of ionic separation $r_{ion}$ in different systems. The equilibrium average (dashed line) and transition state ensemble average (squares/triangles) are shown alongside the negative of the direct interionic force (solid line). The standard deviation of the solvent force is reported to show the actual fluctuation of the solvent force.}
    \label{fig:ion_force}
\end{figure}

As shown in Fig. \ref{fig:ion_force}, we first confirm the previous finding of Ref. \onlinecite{Geissler1999}  that the opposing long range components of the solvent force in the TSE along the interionic axis (blue squares) are significantly different from the full system equilibrium results (blue dashed line), and their ensemble average can cancel most effects of the direct interionic interaction (blue solid line).

Moreover, Fig. \ref{fig:ion_force} shows that in the complete absence of long range Coulomb interactions, the direct interionic force in the CGT system is shifted by $\sim8.5~k_BT/\AA$ relative to the full system, while the solvent force (in both equilibrium and the TSE) is shifted by a similar amount in the opposite direction, such that the effective net force that controls the relative motion between the ion pair is approximately unchanged when we adopt the simplified CGT model. This is consistent with expectations from LMF theory for the influence of the weak and uniformly slowly varying long ranged components from the monovalent ion pair. As a result, the balance between the solvent force and the direct interionic force in the TSE still holds for the CGT system.

\section{Conclusion}

In this work we have explored the question of whether long range forces matter for the transition state and dynamics of NaCl ion-pair dissociation in water. For this purpose, here we have employed Local Molecular Field (LMF) based framework to partition interactions into short range and long range components. Our results illustrate that the short range interactions dominate the dissociation process, and they alone are enough to reproduce the transition rate for the dissociation process, the dynamics of the underlying metastable states and the transition state ensemble.

Overall, in this work the scheme of separating short and long range interaction enabled us to better understand the underlying mechanism of dissociation dynamics in NaCl. While NaCl is an important system routinely studied by many workers, as we show here it is still quite simple in the sense of the minor role of the long ranged interactions. We do however think that the combination of LMF and SPIB should be easily applicable in future work to much more complicated systems. We are particularly interested in investigating systems with multivalent ions, such as CaCl$_2$, where the LMF corrections due to stronger Coulomb tails may have more prominent influences. The LMF-based model reduces the number of sites interacting through long range interactions to that of only charged solute sites, thereby scaling almost linearly with the size of the simulation cell. This feature makes the LMF-based model very compatible with the advanced sampling methods which handle short range interactions efficiently to further speed up the simulation. More interestingly, the combination of the LMF-based models with the AI-based advanced sampling methods, such as RAVE or SPIB,\cite{rave,prave,SPIB} may further facilitate the determination of optimal short range reaction coordinates (RCs) and generate some new insights into the specific physical problems of interest.
\newline

\textbf{Supplementary material\newline }
See supplementary material for details about the molecular dynamics simulation, the implementation of SPIB, the definition of input order parameters, and further analyses of kinetic data.
\newline

\textbf{Acknowledgements\newline }
 This research was entirely supported by the U.S. Department of Energy, Office of Science, Basic Energy Sciences, CPIMS Program, under Award DE-SC0021009. The authors thank Eric Beyerle for proof-reading
the manuscript. This work used the Extreme Science and Engineering Discovery Environment (XSEDE) Bridges through allocation Grant No. CHE180027P, which is supported by the National Science Foundation Grant No. ACI-1548562. We also thank MARCC’s Bluecrab HPC clusters for computing resources. \newline

\textbf{Data availability statement\newline }
The data that support the findings of this study are available from the corresponding author upon reasonable request. \newline

\normalem


\textbf{References}
\begin{thebibliography}{37}%
\makeatletter
\providecommand \@ifxundefined [1]{%
 \@ifx{#1\undefined}
}%
\providecommand \@ifnum [1]{%
 \ifnum #1\expandafter \@firstoftwo
 \else \expandafter \@secondoftwo
 \fi
}%
\providecommand \@ifx [1]{%
 \ifx #1\expandafter \@firstoftwo
 \else \expandafter \@secondoftwo
 \fi
}%
\providecommand \natexlab [1]{#1}%
\providecommand \enquote  [1]{``#1''}%
\providecommand \bibnamefont  [1]{#1}%
\providecommand \bibfnamefont [1]{#1}%
\providecommand \citenamefont [1]{#1}%
\providecommand \href@noop [0]{\@secondoftwo}%
\providecommand \href [0]{\begingroup \@sanitize@url \@href}%
\providecommand \@href[1]{\@@startlink{#1}\@@href}%
\providecommand \@@href[1]{\endgroup#1\@@endlink}%
\providecommand \@sanitize@url [0]{\catcode `\\12\catcode `\$12\catcode
  `\&12\catcode `\#12\catcode `\^12\catcode `\_12\catcode `\%12\relax}%
\providecommand \@@startlink[1]{}%
\providecommand \@@endlink[0]{}%
\providecommand \url  [0]{\begingroup\@sanitize@url \@url }%
\providecommand \@url [1]{\endgroup\@href {#1}{\urlprefix }}%
\providecommand \urlprefix  [0]{URL }%
\providecommand \Eprint [0]{\href }%
\providecommand \doibase [0]{http://dx.doi.org/}%
\providecommand \selectlanguage [0]{\@gobble}%
\providecommand \bibinfo  [0]{\@secondoftwo}%
\providecommand \bibfield  [0]{\@secondoftwo}%
\providecommand \translation [1]{[#1]}%
\providecommand \BibitemOpen [0]{}%
\providecommand \bibitemStop [0]{}%
\providecommand \bibitemNoStop [0]{.\EOS\space}%
\providecommand \EOS [0]{\spacefactor3000\relax}%
\providecommand \BibitemShut  [1]{\csname bibitem#1\endcsname}%
\let\auto@bib@innerbib\@empty
\bibitem [{\citenamefont {Levy}\ and\ \citenamefont
  {Onuchic}(2006)}]{levy2006water}%
  \BibitemOpen
  \bibfield  {author} {\bibinfo {author} {\bibfnamefont {Y.}~\bibnamefont
  {Levy}}\ and\ \bibinfo {author} {\bibfnamefont {J.~N.}\ \bibnamefont
  {Onuchic}},\ }\href@noop {} {\bibfield  {journal} {\bibinfo  {journal} {Annu.
  Rev. Biophys. Biomol. Struct.}\ }\textbf {\bibinfo {volume} {35}},\ \bibinfo
  {pages} {389} (\bibinfo {year} {2006})}\BibitemShut {NoStop}%
\bibitem [{\citenamefont {Dill}(1990)}]{dill1990dominant}%
  \BibitemOpen
  \bibfield  {author} {\bibinfo {author} {\bibfnamefont {K.~A.}\ \bibnamefont
  {Dill}},\ }\href@noop {} {\bibfield  {journal} {\bibinfo  {journal}
  {Biochemistry}\ }\textbf {\bibinfo {volume} {29}},\ \bibinfo {pages} {7133}
  (\bibinfo {year} {1990})}\BibitemShut {NoStop}%
\bibitem [{\citenamefont {Venkateshwaran}, \citenamefont {Vembanur},\ and\
  \citenamefont {Garde}(2014)}]{venkateshwaran2014water}%
  \BibitemOpen
  \bibfield  {author} {\bibinfo {author} {\bibfnamefont {V.}~\bibnamefont
  {Venkateshwaran}}, \bibinfo {author} {\bibfnamefont {S.}~\bibnamefont
  {Vembanur}}, \ and\ \bibinfo {author} {\bibfnamefont {S.}~\bibnamefont
  {Garde}},\ }\href@noop {} {\bibfield  {journal} {\bibinfo  {journal}
  {Proceedings of the National Academy of Sciences}\ }\textbf {\bibinfo
  {volume} {111}},\ \bibinfo {pages} {8729} (\bibinfo {year}
  {2014})}\BibitemShut {NoStop}%
\bibitem [{\citenamefont {Fang}\ \emph {et~al.}(2016)\citenamefont {Fang},
  \citenamefont {Ko}, \citenamefont {Yang}, \citenamefont {Lu},\ and\
  \citenamefont {Hwang}}]{fang2016nucleation}%
  \BibitemOpen
  \bibfield  {author} {\bibinfo {author} {\bibfnamefont {C.-K.}\ \bibnamefont
  {Fang}}, \bibinfo {author} {\bibfnamefont {H.-C.}\ \bibnamefont {Ko}},
  \bibinfo {author} {\bibfnamefont {C.-W.}\ \bibnamefont {Yang}}, \bibinfo
  {author} {\bibfnamefont {Y.-H.}\ \bibnamefont {Lu}}, \ and\ \bibinfo {author}
  {\bibfnamefont {S.}~\bibnamefont {Hwang}},\ }\href@noop {} {\bibfield
  {journal} {\bibinfo  {journal} {Scientific reports}\ }\textbf {\bibinfo
  {volume} {6}},\ \bibinfo {pages} {1} (\bibinfo {year} {2016})}\BibitemShut
  {NoStop}%
\bibitem [{\citenamefont {Allen}\ and\ \citenamefont
  {Tildesley}(2017)}]{allen2017computer}%
  \BibitemOpen
  \bibfield  {author} {\bibinfo {author} {\bibfnamefont {M.~P.}\ \bibnamefont
  {Allen}}\ and\ \bibinfo {author} {\bibfnamefont {D.~J.}\ \bibnamefont
  {Tildesley}},\ }\href@noop {} {\emph {\bibinfo {title} {Computer simulation
  of liquids}}}\ (\bibinfo  {publisher} {Oxford university press},\ \bibinfo
  {year} {2017})\BibitemShut {NoStop}%
\bibitem [{\citenamefont {Belhadj}, \citenamefont {Alper},\ and\ \citenamefont
  {Levy}(1991)}]{belhadj1991molecular}%
  \BibitemOpen
  \bibfield  {author} {\bibinfo {author} {\bibfnamefont {M.}~\bibnamefont
  {Belhadj}}, \bibinfo {author} {\bibfnamefont {H.~E.}\ \bibnamefont {Alper}},
  \ and\ \bibinfo {author} {\bibfnamefont {R.~M.}\ \bibnamefont {Levy}},\
  }\href@noop {} {\bibfield  {journal} {\bibinfo  {journal} {Chemical Physics
  Letters}\ }\textbf {\bibinfo {volume} {179}},\ \bibinfo {pages} {13}
  (\bibinfo {year} {1991})}\BibitemShut {NoStop}%
\bibitem [{\citenamefont {Essmann}\ \emph {et~al.}(1995)\citenamefont
  {Essmann}, \citenamefont {Perera}, \citenamefont {Berkowitz}, \citenamefont
  {Darden}, \citenamefont {Lee},\ and\ \citenamefont
  {Pedersen}}]{essmann1995smooth}%
  \BibitemOpen
  \bibfield  {author} {\bibinfo {author} {\bibfnamefont {U.}~\bibnamefont
  {Essmann}}, \bibinfo {author} {\bibfnamefont {L.}~\bibnamefont {Perera}},
  \bibinfo {author} {\bibfnamefont {M.~L.}\ \bibnamefont {Berkowitz}}, \bibinfo
  {author} {\bibfnamefont {T.}~\bibnamefont {Darden}}, \bibinfo {author}
  {\bibfnamefont {H.}~\bibnamefont {Lee}}, \ and\ \bibinfo {author}
  {\bibfnamefont {L.~G.}\ \bibnamefont {Pedersen}},\ }\href@noop {} {\bibfield
  {journal} {\bibinfo  {journal} {The Journal of chemical physics}\ }\textbf
  {\bibinfo {volume} {103}},\ \bibinfo {pages} {8577} (\bibinfo {year}
  {1995})}\BibitemShut {NoStop}%
\bibitem [{\citenamefont {Schulz}\ \emph {et~al.}(2009)\citenamefont {Schulz},
  \citenamefont {Lindner}, \citenamefont {Petridis},\ and\ \citenamefont
  {Smith}}]{schulz2009scaling}%
  \BibitemOpen
  \bibfield  {author} {\bibinfo {author} {\bibfnamefont {R.}~\bibnamefont
  {Schulz}}, \bibinfo {author} {\bibfnamefont {B.}~\bibnamefont {Lindner}},
  \bibinfo {author} {\bibfnamefont {L.}~\bibnamefont {Petridis}}, \ and\
  \bibinfo {author} {\bibfnamefont {J.~C.}\ \bibnamefont {Smith}},\ }\href@noop
  {} {\bibfield  {journal} {\bibinfo  {journal} {Journal of Chemical Theory and
  Computation}\ }\textbf {\bibinfo {volume} {5}},\ \bibinfo {pages} {2798}
  (\bibinfo {year} {2009})}\BibitemShut {NoStop}%
\bibitem [{\citenamefont {Sun}\ \emph {et~al.}(2012)\citenamefont {Sun},
  \citenamefont {Zheng}, \citenamefont {Mei}, \citenamefont {Bohm},
  \citenamefont {Phillips}, \citenamefont {Kal{\'e}},\ and\ \citenamefont
  {Jones}}]{sun2012optimizing}%
  \BibitemOpen
  \bibfield  {author} {\bibinfo {author} {\bibfnamefont {Y.}~\bibnamefont
  {Sun}}, \bibinfo {author} {\bibfnamefont {G.}~\bibnamefont {Zheng}}, \bibinfo
  {author} {\bibfnamefont {C.}~\bibnamefont {Mei}}, \bibinfo {author}
  {\bibfnamefont {E.~J.}\ \bibnamefont {Bohm}}, \bibinfo {author}
  {\bibfnamefont {J.~C.}\ \bibnamefont {Phillips}}, \bibinfo {author}
  {\bibfnamefont {L.~V.}\ \bibnamefont {Kal{\'e}}}, \ and\ \bibinfo {author}
  {\bibfnamefont {T.~R.}\ \bibnamefont {Jones}},\ }in\ \href@noop {} {\emph
  {\bibinfo {booktitle} {SC'12: Proceedings of the International Conference on
  High Performance Computing, Networking, Storage and Analysis}}}\ (\bibinfo
  {organization} {IEEE},\ \bibinfo {year} {2012})\ pp.\ \bibinfo {pages}
  {1--11}\BibitemShut {NoStop}%
\bibitem [{\citenamefont {Kutzner}\ \emph {et~al.}(2014)\citenamefont
  {Kutzner}, \citenamefont {Apostolov}, \citenamefont {Hess},\ and\
  \citenamefont {Grubm{\"u}ller}}]{kutzner2014scaling}%
  \BibitemOpen
  \bibfield  {author} {\bibinfo {author} {\bibfnamefont {C.}~\bibnamefont
  {Kutzner}}, \bibinfo {author} {\bibfnamefont {R.}~\bibnamefont {Apostolov}},
  \bibinfo {author} {\bibfnamefont {B.}~\bibnamefont {Hess}}, \ and\ \bibinfo
  {author} {\bibfnamefont {H.}~\bibnamefont {Grubm{\"u}ller}},\ }in\ \href@noop
  {} {\emph {\bibinfo {booktitle} {Parallel Computing: Accelerating
  Computational Science and Engineering (CSE)}}}\ (\bibinfo  {publisher} {IOS
  Press},\ \bibinfo {year} {2014})\ pp.\ \bibinfo {pages}
  {722--730}\BibitemShut {NoStop}%
\bibitem [{\citenamefont {Smith}\ and\ \citenamefont {Dang}(1994)}]{Smith1994}%
  \BibitemOpen
  \bibfield  {author} {\bibinfo {author} {\bibfnamefont {D.~E.}\ \bibnamefont
  {Smith}}\ and\ \bibinfo {author} {\bibfnamefont {L.~X.}\ \bibnamefont
  {Dang}},\ }\href@noop {} {\bibfield  {journal} {\bibinfo  {journal} {The
  Journal of Chemical Physics}\ }\textbf {\bibinfo {volume} {100}},\ \bibinfo
  {pages} {3757} (\bibinfo {year} {1994})}\BibitemShut {NoStop}%
\bibitem [{\citenamefont {Geissler}, \citenamefont {Dellago},\ and\
  \citenamefont {Chandler}(1999)}]{Geissler1999}%
  \BibitemOpen
  \bibfield  {author} {\bibinfo {author} {\bibfnamefont {P.~L.}\ \bibnamefont
  {Geissler}}, \bibinfo {author} {\bibfnamefont {C.}~\bibnamefont {Dellago}}, \
  and\ \bibinfo {author} {\bibfnamefont {D.}~\bibnamefont {Chandler}},\
  }\href@noop {} {\bibfield  {journal} {\bibinfo  {journal} {The Journal of
  Physical Chemistry B}\ }\textbf {\bibinfo {volume} {103}},\ \bibinfo {pages}
  {3706} (\bibinfo {year} {1999})}\BibitemShut {NoStop}%
\bibitem [{\citenamefont {Ballard}\ and\ \citenamefont
  {Dellago}(2012)}]{Ballard2012}%
  \BibitemOpen
  \bibfield  {author} {\bibinfo {author} {\bibfnamefont {A.~J.}\ \bibnamefont
  {Ballard}}\ and\ \bibinfo {author} {\bibfnamefont {C.}~\bibnamefont
  {Dellago}},\ }\href@noop {} {\bibfield  {journal} {\bibinfo  {journal} {The
  Journal of Physical Chemistry B}\ }\textbf {\bibinfo {volume} {116}},\
  \bibinfo {pages} {13490} (\bibinfo {year} {2012})}\BibitemShut {NoStop}%
\bibitem [{\citenamefont {Mullen}, \citenamefont {Shea},\ and\ \citenamefont
  {Peters}(2014)}]{Mullen2014}%
  \BibitemOpen
  \bibfield  {author} {\bibinfo {author} {\bibfnamefont {R.~G.}\ \bibnamefont
  {Mullen}}, \bibinfo {author} {\bibfnamefont {J.-E.}\ \bibnamefont {Shea}}, \
  and\ \bibinfo {author} {\bibfnamefont {B.}~\bibnamefont {Peters}},\
  }\href@noop {} {\bibfield  {journal} {\bibinfo  {journal} {Journal of
  chemical theory and computation}\ }\textbf {\bibinfo {volume} {10}},\
  \bibinfo {pages} {659} (\bibinfo {year} {2014})}\BibitemShut {NoStop}%
\bibitem [{\citenamefont {Bolhuis}\ \emph {et~al.}(2002)\citenamefont
  {Bolhuis}, \citenamefont {Chandler}, \citenamefont {Dellago},\ and\
  \citenamefont {Geissler}}]{committor_review}%
  \BibitemOpen
  \bibfield  {author} {\bibinfo {author} {\bibfnamefont {P.~G.}\ \bibnamefont
  {Bolhuis}}, \bibinfo {author} {\bibfnamefont {D.}~\bibnamefont {Chandler}},
  \bibinfo {author} {\bibfnamefont {C.}~\bibnamefont {Dellago}}, \ and\
  \bibinfo {author} {\bibfnamefont {P.~L.}\ \bibnamefont {Geissler}},\
  }\href@noop {} {\bibfield  {journal} {\bibinfo  {journal} {Advances in
  chemical physics}\ }\textbf {\bibinfo {volume} {53}},\ \bibinfo {pages} {291}
  (\bibinfo {year} {2002})}\BibitemShut {NoStop}%
\bibitem [{\citenamefont {Chen}, \citenamefont {Kaur},\ and\ \citenamefont
  {Weeks}(2004)}]{chen2004connecting}%
  \BibitemOpen
  \bibfield  {author} {\bibinfo {author} {\bibfnamefont {Y.-g.}\ \bibnamefont
  {Chen}}, \bibinfo {author} {\bibfnamefont {C.}~\bibnamefont {Kaur}}, \ and\
  \bibinfo {author} {\bibfnamefont {J.~D.}\ \bibnamefont {Weeks}},\ }\href@noop
  {} {\bibfield  {journal} {\bibinfo  {journal} {The Journal of Physical
  Chemistry B}\ }\textbf {\bibinfo {volume} {108}},\ \bibinfo {pages} {19874}
  (\bibinfo {year} {2004})}\BibitemShut {NoStop}%
\bibitem [{\citenamefont {Chen}\ and\ \citenamefont
  {Weeks}(2006)}]{chen2006local}%
  \BibitemOpen
  \bibfield  {author} {\bibinfo {author} {\bibfnamefont {Y.-G.}\ \bibnamefont
  {Chen}}\ and\ \bibinfo {author} {\bibfnamefont {J.~D.}\ \bibnamefont
  {Weeks}},\ }\href@noop {} {\bibfield  {journal} {\bibinfo  {journal}
  {Proceedings of the National Academy of Sciences}\ }\textbf {\bibinfo
  {volume} {103}},\ \bibinfo {pages} {7560} (\bibinfo {year}
  {2006})}\BibitemShut {NoStop}%
\bibitem [{\citenamefont {Rodgers}\ and\ \citenamefont
  {Weeks}(2008{\natexlab{a}})}]{rodgers2008interplay}%
  \BibitemOpen
  \bibfield  {author} {\bibinfo {author} {\bibfnamefont {J.~M.}\ \bibnamefont
  {Rodgers}}\ and\ \bibinfo {author} {\bibfnamefont {J.~D.}\ \bibnamefont
  {Weeks}},\ }\href@noop {} {\bibfield  {journal} {\bibinfo  {journal}
  {Proceedings of the National Academy of Sciences}\ }\textbf {\bibinfo
  {volume} {105}},\ \bibinfo {pages} {19136} (\bibinfo {year}
  {2008}{\natexlab{a}})}\BibitemShut {NoStop}%
\bibitem [{\citenamefont {Rodgers}\ and\ \citenamefont
  {Weeks}(2008{\natexlab{b}})}]{rodgers2008local}%
  \BibitemOpen
  \bibfield  {author} {\bibinfo {author} {\bibfnamefont {J.~M.}\ \bibnamefont
  {Rodgers}}\ and\ \bibinfo {author} {\bibfnamefont {J.~D.}\ \bibnamefont
  {Weeks}},\ }\href@noop {} {\bibfield  {journal} {\bibinfo  {journal} {Journal
  of Physics: Condensed Matter}\ }\textbf {\bibinfo {volume} {20}},\ \bibinfo
  {pages} {494206} (\bibinfo {year} {2008}{\natexlab{b}})}\BibitemShut
  {NoStop}%
\bibitem [{\citenamefont {Weeks}(2002)}]{weeks2002connecting}%
  \BibitemOpen
  \bibfield  {author} {\bibinfo {author} {\bibfnamefont {J.~D.}\ \bibnamefont
  {Weeks}},\ }\href@noop {} {\bibfield  {journal} {\bibinfo  {journal} {Annual
  review of physical chemistry}\ }\textbf {\bibinfo {volume} {53}},\ \bibinfo
  {pages} {533} (\bibinfo {year} {2002})}\BibitemShut {NoStop}%
\bibitem [{\citenamefont {Weeks}, \citenamefont {Chandler},\ and\ \citenamefont
  {Andersen}(1971)}]{weeks1971role}%
  \BibitemOpen
  \bibfield  {author} {\bibinfo {author} {\bibfnamefont {J.~D.}\ \bibnamefont
  {Weeks}}, \bibinfo {author} {\bibfnamefont {D.}~\bibnamefont {Chandler}}, \
  and\ \bibinfo {author} {\bibfnamefont {H.~C.}\ \bibnamefont {Andersen}},\
  }\href@noop {} {\bibfield  {journal} {\bibinfo  {journal} {The Journal of
  chemical physics}\ }\textbf {\bibinfo {volume} {54}},\ \bibinfo {pages}
  {5237} (\bibinfo {year} {1971})}\BibitemShut {NoStop}%
\bibitem [{\citenamefont {Wang}\ and\ \citenamefont {Tiwary}(2021)}]{SPIB}%
  \BibitemOpen
  \bibfield  {author} {\bibinfo {author} {\bibfnamefont {D.}~\bibnamefont
  {Wang}}\ and\ \bibinfo {author} {\bibfnamefont {P.}~\bibnamefont {Tiwary}},\
  }\href@noop {} {\bibfield  {journal} {\bibinfo  {journal} {The Journal of
  Chemical Physics}\ }\textbf {\bibinfo {volume} {154}},\ \bibinfo {pages}
  {134111} (\bibinfo {year} {2021})}\BibitemShut {NoStop}%
\bibitem [{\citenamefont {Remsing}, \citenamefont {Rodgers},\ and\
  \citenamefont {Weeks}(2011)}]{remsing2011deconstructing}%
  \BibitemOpen
  \bibfield  {author} {\bibinfo {author} {\bibfnamefont {R.~C.}\ \bibnamefont
  {Remsing}}, \bibinfo {author} {\bibfnamefont {J.~M.}\ \bibnamefont
  {Rodgers}}, \ and\ \bibinfo {author} {\bibfnamefont {J.~D.}\ \bibnamefont
  {Weeks}},\ }\href@noop {} {\bibfield  {journal} {\bibinfo  {journal} {Journal
  of Statistical Physics}\ }\textbf {\bibinfo {volume} {145}},\ \bibinfo
  {pages} {313} (\bibinfo {year} {2011})}\BibitemShut {NoStop}%
\bibitem [{\citenamefont {Gao}, \citenamefont {Remsing},\ and\ \citenamefont
  {Weeks}(2020)}]{gao2020short}%
  \BibitemOpen
  \bibfield  {author} {\bibinfo {author} {\bibfnamefont {A.}~\bibnamefont
  {Gao}}, \bibinfo {author} {\bibfnamefont {R.~C.}\ \bibnamefont {Remsing}}, \
  and\ \bibinfo {author} {\bibfnamefont {J.~D.}\ \bibnamefont {Weeks}},\
  }\href@noop {} {\bibfield  {journal} {\bibinfo  {journal} {Proceedings of the
  National Academy of Sciences}\ }\textbf {\bibinfo {volume} {117}},\ \bibinfo
  {pages} {1293} (\bibinfo {year} {2020})}\BibitemShut {NoStop}%
\bibitem [{\citenamefont {Wang}, \citenamefont {Ribeiro},\ and\ \citenamefont
  {Tiwary}(2019)}]{prave}%
  \BibitemOpen
  \bibfield  {author} {\bibinfo {author} {\bibfnamefont {Y.}~\bibnamefont
  {Wang}}, \bibinfo {author} {\bibfnamefont {J.~M.~L.}\ \bibnamefont
  {Ribeiro}}, \ and\ \bibinfo {author} {\bibfnamefont {P.}~\bibnamefont
  {Tiwary}},\ }\href {\doibase 10.1038/s41467-019-11405-4} {\bibfield
  {journal} {\bibinfo  {journal} {Nat. Commun.}\ }\textbf {\bibinfo {volume}
  {10}},\ \bibinfo {pages} {3573} (\bibinfo {year} {2019})}\BibitemShut
  {NoStop}%
\bibitem [{\citenamefont {Chandler}, \citenamefont {Weeks},\ and\ \citenamefont
  {Andersen}(1983)}]{chandler1983van}%
  \BibitemOpen
  \bibfield  {author} {\bibinfo {author} {\bibfnamefont {D.}~\bibnamefont
  {Chandler}}, \bibinfo {author} {\bibfnamefont {J.~D.}\ \bibnamefont {Weeks}},
  \ and\ \bibinfo {author} {\bibfnamefont {H.~C.}\ \bibnamefont {Andersen}},\
  }\href@noop {} {\bibfield  {journal} {\bibinfo  {journal} {Science}\ }\textbf
  {\bibinfo {volume} {220}},\ \bibinfo {pages} {787} (\bibinfo {year}
  {1983})}\BibitemShut {NoStop}%
\bibitem [{\citenamefont {Remsing}\ \emph {et~al.}(2018)\citenamefont
  {Remsing}, \citenamefont {Duignan}, \citenamefont {Baer}, \citenamefont
  {Schenter}, \citenamefont {Mundy},\ and\ \citenamefont
  {Weeks}}]{remsing2018water}%
  \BibitemOpen
  \bibfield  {author} {\bibinfo {author} {\bibfnamefont {R.~C.}\ \bibnamefont
  {Remsing}}, \bibinfo {author} {\bibfnamefont {T.~T.}\ \bibnamefont
  {Duignan}}, \bibinfo {author} {\bibfnamefont {M.~D.}\ \bibnamefont {Baer}},
  \bibinfo {author} {\bibfnamefont {G.~K.}\ \bibnamefont {Schenter}}, \bibinfo
  {author} {\bibfnamefont {C.~J.}\ \bibnamefont {Mundy}}, \ and\ \bibinfo
  {author} {\bibfnamefont {J.~D.}\ \bibnamefont {Weeks}},\ }\href@noop {}
  {\bibfield  {journal} {\bibinfo  {journal} {The Journal of Physical Chemistry
  B}\ }\textbf {\bibinfo {volume} {122}},\ \bibinfo {pages} {3519} (\bibinfo
  {year} {2018})}\BibitemShut {NoStop}%
\bibitem [{\citenamefont {Ribeiro}\ \emph {et~al.}(2018)\citenamefont
  {Ribeiro}, \citenamefont {Bravo}, \citenamefont {Wang},\ and\ \citenamefont
  {Tiwary}}]{rave}%
  \BibitemOpen
  \bibfield  {author} {\bibinfo {author} {\bibfnamefont {J.~M.~L.}\
  \bibnamefont {Ribeiro}}, \bibinfo {author} {\bibfnamefont {P.}~\bibnamefont
  {Bravo}}, \bibinfo {author} {\bibfnamefont {Y.}~\bibnamefont {Wang}}, \ and\
  \bibinfo {author} {\bibfnamefont {P.}~\bibnamefont {Tiwary}},\ }\href
  {\doibase 10.1063/1.5025487} {\bibfield  {journal} {\bibinfo  {journal} {J.
  Chem. Phys.}\ }\textbf {\bibinfo {volume} {149}},\ \bibinfo {pages} {072301}
  (\bibinfo {year} {2018})}\BibitemShut {NoStop}%
\bibitem [{\citenamefont {Alemi}\ \emph {et~al.}(2016)\citenamefont {Alemi},
  \citenamefont {Fischer}, \citenamefont {Dillon},\ and\ \citenamefont
  {Murphy}}]{variational_IB}%
  \BibitemOpen
  \bibfield  {author} {\bibinfo {author} {\bibfnamefont {A.~A.}\ \bibnamefont
  {Alemi}}, \bibinfo {author} {\bibfnamefont {I.}~\bibnamefont {Fischer}},
  \bibinfo {author} {\bibfnamefont {J.~V.}\ \bibnamefont {Dillon}}, \ and\
  \bibinfo {author} {\bibfnamefont {K.}~\bibnamefont {Murphy}},\ }\href@noop {}
  {\bibfield  {journal} {\bibinfo  {journal} {arXiv preprint arXiv:1612.00410}\
  } (\bibinfo {year} {2016})}\BibitemShut {NoStop}%
\bibitem [{\citenamefont {Lorentz}(1881)}]{lorentz1881ueber}%
  \BibitemOpen
  \bibfield  {author} {\bibinfo {author} {\bibfnamefont {H.~A.}\ \bibnamefont
  {Lorentz}},\ }\href@noop {} {\bibfield  {journal} {\bibinfo  {journal}
  {Annalen der physik}\ }\textbf {\bibinfo {volume} {248}},\ \bibinfo {pages}
  {127} (\bibinfo {year} {1881})}\BibitemShut {NoStop}%
\bibitem [{\citenamefont {Plimpton}(1995)}]{plimpton1995fast}%
  \BibitemOpen
  \bibfield  {author} {\bibinfo {author} {\bibfnamefont {S.}~\bibnamefont
  {Plimpton}},\ }\href@noop {} {\bibfield  {journal} {\bibinfo  {journal}
  {Journal of computational physics}\ }\textbf {\bibinfo {volume} {117}},\
  \bibinfo {pages} {1} (\bibinfo {year} {1995})}\BibitemShut {NoStop}%
\bibitem [{\citenamefont {Nos{\'e}}(1984)}]{nose1984unified}%
  \BibitemOpen
  \bibfield  {author} {\bibinfo {author} {\bibfnamefont {S.}~\bibnamefont
  {Nos{\'e}}},\ }\href@noop {} {\bibfield  {journal} {\bibinfo  {journal} {The
  Journal of chemical physics}\ }\textbf {\bibinfo {volume} {81}},\ \bibinfo
  {pages} {511} (\bibinfo {year} {1984})}\BibitemShut {NoStop}%
\bibitem [{\citenamefont {Nos{\'e}}\ and\ \citenamefont
  {Klein}(1983)}]{nose1983constant}%
  \BibitemOpen
  \bibfield  {author} {\bibinfo {author} {\bibfnamefont {S.}~\bibnamefont
  {Nos{\'e}}}\ and\ \bibinfo {author} {\bibfnamefont {M.}~\bibnamefont
  {Klein}},\ }\href@noop {} {\bibfield  {journal} {\bibinfo  {journal}
  {Molecular Physics}\ }\textbf {\bibinfo {volume} {50}},\ \bibinfo {pages}
  {1055} (\bibinfo {year} {1983})}\BibitemShut {NoStop}%
\bibitem [{\citenamefont {Salvalaglio}, \citenamefont {Tiwary},\ and\
  \citenamefont {Parrinello}(2014)}]{Salvalaglio2014}%
  \BibitemOpen
  \bibfield  {author} {\bibinfo {author} {\bibfnamefont {M.}~\bibnamefont
  {Salvalaglio}}, \bibinfo {author} {\bibfnamefont {P.}~\bibnamefont {Tiwary}},
  \ and\ \bibinfo {author} {\bibfnamefont {M.}~\bibnamefont {Parrinello}},\
  }\href@noop {} {\bibfield  {journal} {\bibinfo  {journal} {Journal of
  chemical theory and computation}\ }\textbf {\bibinfo {volume} {10}},\
  \bibinfo {pages} {1420} (\bibinfo {year} {2014})}\BibitemShut {NoStop}%
\bibitem [{\citenamefont {Kemp}, \citenamefont {Zaradic},\ and\ \citenamefont
  {Hansen}(2007)}]{Kemp2007}%
  \BibitemOpen
  \bibfield  {author} {\bibinfo {author} {\bibfnamefont {S.~J.}\ \bibnamefont
  {Kemp}}, \bibinfo {author} {\bibfnamefont {P.}~\bibnamefont {Zaradic}}, \
  and\ \bibinfo {author} {\bibfnamefont {F.}~\bibnamefont {Hansen}},\
  }\href@noop {} {\bibfield  {journal} {\bibinfo  {journal} {Ecological
  modelling}\ }\textbf {\bibinfo {volume} {204}},\ \bibinfo {pages} {326}
  (\bibinfo {year} {2007})}\BibitemShut {NoStop}%
\bibitem [{\citenamefont {Jung}, \citenamefont {Covino},\ and\ \citenamefont
  {Hummer}(2019)}]{Hummer2019}%
  \BibitemOpen
  \bibfield  {author} {\bibinfo {author} {\bibfnamefont {H.}~\bibnamefont
  {Jung}}, \bibinfo {author} {\bibfnamefont {R.}~\bibnamefont {Covino}}, \ and\
  \bibinfo {author} {\bibfnamefont {G.}~\bibnamefont {Hummer}},\ }\href@noop {}
  {\bibfield  {journal} {\bibinfo  {journal} {arXiv preprint arXiv:1901.04595}\
  } (\bibinfo {year} {2019})}\BibitemShut {NoStop}%
\bibitem [{\citenamefont {Rey}\ and\ \citenamefont {Guardia}(1992)}]{Rey1992}%
  \BibitemOpen
  \bibfield  {author} {\bibinfo {author} {\bibfnamefont {R.}~\bibnamefont
  {Rey}}\ and\ \bibinfo {author} {\bibfnamefont {E.}~\bibnamefont {Guardia}},\
  }\href@noop {} {\bibfield  {journal} {\bibinfo  {journal} {The Journal of
  Physical Chemistry}\ }\textbf {\bibinfo {volume} {96}},\ \bibinfo {pages}
  {4712} (\bibinfo {year} {1992})}\BibitemShut {NoStop}%
\end{thebibliography}
	\end{document}